\documentclass{elsart}
\usepackage[dvips]{graphicx}
\begin{document}
\begin{frontmatter}
\title{Combinatorial Level Densities from a  
Relativistic Structure Model}
\author[label1]{R. Pezer}
\address[label1]{Physics Department, Faculty of Science, University of Zagreb,
                 Croatia}
\author[label2]{A. Ventura\thanksref{label4}\corauthref{cor1}}
\address[label2]{ Ente Nuove Tecnologie, Energia e Ambiente, Bologna, Italy\\
                 and Istituto Nazionale di Fisica Nucleare,
                 Sezione di Bologna, Italy}
\corauth[cor1]{Corresponding author.}
\thanks[label4]{Address: ENEA - Via Martiri di Monte Sole 4 -
                 40129 Bologna, Italy.}
\ead{ventura@bologna.enea.it}
\author[label3]{D. Vretenar}
\address[label3]{Physics Department, Faculty of Science, University of Zagreb,
                Croatia}

\begin{abstract}
A new model for calculating nuclear level
densities is investigated. The single-nucleon spectra are
calculated in a relativistic mean-field model
with energy-dependent effective mass, 
which yields a realistic density of single-particle states at the Fermi
energy. These microscopic single-nucleon states are used in a fast 
combinatorial algorithm for calculating the non-collective
excitations of nuclei. The method, when applied to magic and semi-magic
nuclei, such as $^{60}$Ni, $^{114}$Sn and $^{208}$Pb, reproduces the
cumulative number of experimental states at low excitation energy, as well
as the $s$-wave neutron resonance spacing at the neutron binding energy.
Experimental level densities above 10 MeV are reproduced by multiplying the
non-collective level densities by a simple vibrational enhancement factor.
Problems to be solved in the extension to open-shell nuclei are discussed.
\end{abstract}       


\end{frontmatter}

\section{Introduction}
\label{sec:intro}

Nuclear level densities have been a lively field of research for over sixty
years since the pioneering work of Bethe\cite{Bethe:1937:AA}. The subject
is interesting both from a purely theoretical point of view (the problem 
of a quantum many-body system with continuum excitation energy), as well as
from a perspective of applications (e.g. an essential ingredient of statistical
models of nuclear reactions).

Owing to their computational simplicity, phenomenological formulae 
for nuclear level densities with
parameters adjusted to empirical data available at low
excitation energy (cumulative numbers of discrete levels and
s-wave neutron resonance spacings), have been used for many years. 
Their extrapolation to those regions of the periodic chart where 
there is no experimental information is, however, doubtful. 
The calculation of nuclear level densities should be rather based
on microscopic structure models, and should make use of 
fast computational algorithms in order to deal with huge numbers
of states that increase exponentially with excitation 
energy \cite{Bethe:1937:AA}. 

In this work we investigate a new model for calculating nuclear level
densities. The method combines an efficient combinatorial 
algorithm based on the
Gaussian polynomial expansion of a generating function~\cite{Sunko:1997:AA}, 
and an improved relativistic mean-field structure model
with energy-dependent effective mass~\cite{Vretenar:2002:PDN}. The microscopic
structure model not only reproduces the bulk nuclear properties over 
the whole periodic chart, but it also yields
realistic densities of single-nucleon states around
the Fermi energy, an essential requirement for a self-consistent
mean-field approach to nuclear level densities.

In Section \ref{sec:rsm} we outline the relativistic mean-field model
with energy-dependent effective mass.
The combinatorial algorithm for calculating nuclear level densities 
is described in Section \ref{sec:comb-algo}. In this work applications 
are limited to magic and semi-magic nuclei. Empirical data at
excitation energies above 10 MeV can be reproduced at the price of
introducing a vibrational enhancement factor, described in 
Section \ref{sec:coll-eff}. Our numerical results are 
compared with experimental data in Section \ref{sec:num-res}.
In Sec.~\ref{sec:concl} we summarize the results 
and present an outlook for future applications.

\section{Relativistic Structure Model}
\label{sec:rsm}

The microscopic nuclear structure model adopted in the present
work is described in detail in
Ref. \cite{Vretenar:2002:PDN}, where it has also 
been applied in the calculation of single-nucleon
states in magic nuclei, e.g. $^{132}$Sn and $^{208}$Pb. 
In order to make the present analysis self-contained,
the essential features of the model are summarized in this section.
Here and in the following, use is made of units $\hbar =c =1$.

In the relativistic mean-field approximation \cite{Serot:1986:PDN},
nucleons are described as point particles that
move independently in mean fields
which originate from the nucleon-nucleon interaction.
The theory is fully Lorentz invariant.
Conditions of causality and Lorentz invariance impose that the
interaction is mediated by the
exchange of point-like effective mesons, which couple to the nucleons
at local vertices. The single-nucleon dynamics is described by the
Dirac equation
\begin{equation}
\label{eq:statDirac}
\left\{-i\mbox{\boldmath $\alpha$}
\cdot\mbox{\boldmath $\nabla$}
+\beta(m+g_\sigma \sigma)
+g_\omega \omega^0+g_\rho\tau_3\rho^0_3
+e\frac{(1-\tau_3)}{2} A^0\right\}\psi_i=
\varepsilon_i\psi_i.
\end{equation}
$\sigma$, $\omega$, and
$\rho$ are the meson fields, and $A$ denotes the electromagnetic potential.
$g_\sigma$ $g_\omega$, and $g_\rho$ are the corresponding coupling
constants for the mesons to the nucleon.
The lowest order of the quantum field theory is the {\it
mean-field} approximation: the meson field operators are
replaced by their expectation values. The sources
of the meson fields are defined by the nucleon densities
and currents.  The ground state of a nucleus is described
by the stationary self-consistent solution of the coupled
system of the Dirac~(\ref{eq:statDirac}) and Klein-Gordon equations:
\begin{eqnarray}
\left[ -\Delta +m_{\sigma }^{2}\right] \,\sigma ({\bf r}) &=&-g_{\sigma
}\,\rho _{s}({\bf r})-g_{2}\,\sigma ^{2}({\bf r})-g_{3}\,\sigma ^{3}({\bf r})
\label{messig} \\
\left[ -\Delta +m_{\omega }^{2}\right] \,\omega ^{0}({\bf r}) &=&g_{\omega
}\,\rho _{v}({\bf r})  
\label{mesome} \\
\left[ -\Delta +m_{\rho }^{2}\right] \,\rho ^{0}({\bf r}) &=&g_{\rho }\,\rho
_{3}({\bf r})  
\label{mesrho} \\
-\Delta \,A^{0}({\bf r}) &=&e\,\rho _{p}({\bf r}),  
\label{photon}
\end{eqnarray}
for the sigma meson, omega meson, rho meson and photon field, respectively.
Due to charge conservation, only the 3rd-component of the isovector 
\mbox{\boldmath $\vec \rho$}
meson contributes. The source terms in equations (\ref{messig}) to (\ref
{photon}) are sums of bilinear products of baryon amplitudes, and they
are calculated in the {\it no-sea} approximation, i.e. the Dirac sea
of negative energy states does not contribute to the nucleon densities
and currents. Due to time reversal invariance,
there are no currents in the static solution for an even-even
system, and therefore the spatial
vector components \mbox{\boldmath $\omega,~\rho_3$} and
${\bf  A}$ of the vector meson fields vanish.
The quartic potential
\begin{equation}
U(\sigma )~=~\frac{1}{2}m_{\sigma }^{2}\sigma ^{2}+\frac{1}{3}g_{2}\sigma
^{3}+\frac{1}{4}g_{3}\sigma ^{4}  \nonumber
\end{equation}
introduces an effective density dependence. The non-linear
self-interaction of the $\sigma$ field is essential for
a quantitative description of properties of finite nuclei.

The effective single-nucleon potential   
is essentially given by the sum of the scalar attractive
$\sigma$-potential and the vector repulsive
$\omega$-potential.
Both potentials are of the order of several hundred MeV 
in the nuclear interior. The contributions of the isovector 
$\rho$-meson field and the electromagnetic interaction are much smaller.
In the relativistic Hartree mean-field approximation the nucleon self-energy 
is real, local and energy independent. However, due to
the momentum dependence of the scalar density or, equivalently, 
the momentum dependence of the Dirac mass in the non-relativistic 
reduction of the Dirac equation, even in the Hartree approximation
the equivalent Schr\" odinger 
potential is nonlocal, i.e. momentum dependent\cite{Vretenar:2002:PDN}. 

In applications of the standard relativistic mean-field model
to the description of ground state properties of spherical
and deformed nuclei, the approximation of the real, local and 
energy independent nucleon self-energy leads to the well known problem of 
low effective mass, i.e. low density of single-nucleon 
states close to the Fermi surface.
The problem of defining, in the relativistic framework, a 
local energy-dependent potential equivalent to a
microscopic non-local potential, was discussed in 
detail in Ref. \cite{Jaminon:1989:AA}. In a subsequent 
article~\cite{Jaminon:1990:AA} the same authors applied the relativistic 
Br\H{u}ckner-Hartree-Fock approximation to the calculation of dispersion
relations that connect the real and imaginary parts of the Lorentz components of
the mean field in symmetric nuclear matter and showed, in particular, that
the total dispersive contribution to the real part of the mean field depends
almost linearly on energy
 in an interval of half-width 10 MeV around
the Fermi energy $E_F$.

The effect of non-locality in space and time of the underlying microscopic
potential on the equivalent local and energy dependent potential in a
relativistic theory, is conveniently expressed in terms of an effective
nucleon mass $m^{*}$. The effective mass should
not to be confused, however, with the Dirac mass, $m_D=m+g_\sigma \sigma $, 
appearing in Eq. (\ref{eq:statDirac}). The latter is always
smaller than $m$, due to the fact that 
$g_\sigma \sigma $ is an attractive potential,
of the order of $0.5m$ in the nuclear interior, and very close to $m$ at the
surface, where $\sigma $ goes to zero. 
As shown in detail in Refs.\cite{Jaminon:1989:AA,Jaminon:1990:AA}, 
the effective mass which is 
directly connected with the density of single particle states around 
the Fermi energy $E_F$, i.e. the crucial parameter of any nuclear
state density model, can be defined as follows:
\begin{equation}
\frac{m^{*}\left( E\right) }m=1-\frac{dV_e\left( E\right) }{dE}\, ,
\label{Effm}
\end{equation}
where $E = \varepsilon - m$ is the difference between the total nucleon
energy, $\varepsilon$, and its rest mass, and
$V_e\left( E\right) $ is the real part of the energy-dependent Schr\H{o%
}dinger-equivalent potential. In symmetric nuclear matter 
$V_e$ reads \cite {Jaminon:1989:AA}: 
\begin{equation}
V_e\left( E\right) =V_\sigma +V_\omega +\frac EmV_\omega +\frac 1{2m}\left(
V_\sigma ^2-V_\omega ^2\right) \, ,  
\label{eq:Ve-nu-mat}
\end{equation}
where $V_{\sigma (\omega) }$ is the real part of the $\sigma (\omega) $
potential.
Eqs. (\ref{Effm}-\ref{eq:Ve-nu-mat}) show that in the standard mean-field approximation,
when $V_\sigma $ and $V_\omega $ do not depend on energy, the effective
nucleon mass is also energy independent. The theoretical
results of Ref. \cite{Jaminon:1990:AA} can be phenomenologically 
reproduced by assuming
that $V_\sigma $ and $V_\omega $ are linear functions of $E$ in an energy
window of half-width, $\Delta E$, of the order of 10 MeV centered at $E_F$.
In addition, we shall make the same simplifying assumption as in Ref. \cite
{Vretenar:2002:PDN}, $dV_\sigma /dE=dV_\omega /dE=\alpha $, or, equivalently 
\begin{equation}
V_{\sigma (\omega) }=V_{\sigma (\omega) }^0+\alpha \left( E-E_F\right),%
\qquad 
\left( \left| E-E_F\right| \leq \Delta E\right)  
\label{eq:VE}
\end{equation}
With this assumption, $m^{*}\left( E\right) $ becomes a linear function
of $E$ in the same energy window: 
\begin{eqnarray}
\frac{m^{*}\left( E\right) }m &=& 
            1-2\alpha -\frac{V_\omega ^0}m
	      \left( 1-\alpha
              \right) -
	      \frac{V_\sigma ^0}m\alpha +
	      \frac{\alpha E_F}m-2\frac{\alpha E}m%
\nonumber \\
 \left| E-E_F\right| &\leq& \Delta E \, . 
\nonumber
\end{eqnarray}
In the case of finite spherical nuclei, considered in Ref. \cite
{Vretenar:2002:PDN} and in the present work, $V_\sigma ^{}$ and $V_\omega ^{}$
are, of course, functions of the radial coordinate, and so are the
Schr\"odinger equivalent potential, which contains additional $r$-derivative
terms with respect to Eq. (\ref{eq:Ve-nu-mat}), and the effective
nucleon mass. However, as opposed to the Dirac mass $m_D$, $m^{*}(r,E)$ might
become greater than $m$, provided $\alpha $ is negative and large enough.

It is also important to point out that the present definition of $m^{*}$ 
can be directly compared to the effective mass in non-relativistic 
models aimed at reproducing the empirical nuclear state densities, such as
the model of Ref. \cite{Shlomo:1991:AA} which 
accounts for the observed decrease of the
level density parameter $a=\frac{\pi ^2}6g\left( E_F\right) $ of the Bethe
formula~\cite{Bethe:1937:AA} with increasing excitation energy or temperature, 
from the average value of $A/8$ MeV$^{-1}$ at $T\le 1$ MeV to $A/11$ MeV$%
^{-1}$ at $T\simeq 2.5$ MeV in the mass region $A\simeq 160$. Since the
single-particle level density at the Fermi energy, $g\left( E_F\right) $, is
empirically proportional to the mass number $A$ and,
in the model of Ref. \cite{Shlomo:1991:AA}, depends almost
linearly on $m^{*}\left( E_F\right) $, we
expect that our $m^{*}\left( E_F\right) $ is a smooth
function of  $A$
in the same mass region. The density-averaged effective mass of Ref. \cite
{Shlomo:1991:AA} ranges from $1.12m$ at $A=40$ to $1.13m$ at $A=210$. 
These values are corroborated by a recent analysis~\cite{Mughabghab:1998:AA} 
of neutron resonance spacings
of several stable nuclei, which yields an average value $\left\langle
m^{*}\left( E_F\right) \right\rangle /m=1.09\pm 0.13$.

If, according to Eq. (\ref{eq:VE}),the effective potential in the Dirac equation 
(\ref{eq:statDirac}) is a linear function of $E$, this defines a generalized
eigenvalue problem 
\begin{equation} \nonumber
\hat{H}_D\psi =\bar{A}E\psi \, ,
\end{equation}
where $\hat{H}_D$ is the energy independent Dirac Hamiltonian and $\bar{A}$ 
denotes the matrix that contains the linear energy
dependence. The problem can be solved exactly; the technical aspects of the
solution are discussed in Ref. \cite{Vretenar:2002:PDN} 
and will not be repeated here.

We conclude this section by noting that, as is well known from
non-relativistic theories, the energy dependence of the nucleon self-energy
can be described  microscopically in terms of the coupling of single-particle
states to surface vibrations (see, e. g., Ref. \cite{Bohr:1975:NS}).
This can be also done in a relativistic framework, by coupling the
single-particle states generated in the energy-independent mean-field
approximation to surface vibrations calculated in the relativistic random
phase approximation. The present phenomenological approach can be considered
as a preliminary quantitative evaluation of the impact of 
particle-vibration coupling on the calculation of total level densities in
spherical nuclei.

\section{Combinatorial Algorithm}
\label{sec:comb-algo}

Several techniques for an exact treatment of the problem of generating 
the full many-body physical state space for simple nuclear models can be
found in the literature. An example is the odometer approach to the
generation of multi-particle configurations adopted by Hillman and Grover 
\cite{Hillman:1969:AA}, which is simple, but not efficient enough to be used in the
case of heavy nuclei at high excitation energy.
Recursive methods of calculating state and level densities of
non-interacting many-fermion systems were proposed by various authors
(~\cite{Williams:1969:AA},~\cite{Albrecht:1973:AA},~\cite{Jacquemin:1986:AA}).
 Berger and Martinot \cite
{Berger:1974:AA} proposed the use of a five-variable generating function, whose
power expansion coefficients determine the number of nuclear states with
given numbers of proton particles, proton holes, neutron particles and
neutron holes at a given excitation energy. This approach has been
generalized by Hilaire, Delaroche and Girod \cite{Hilaire:2001:CNL} and 
applied to the
calculation of nuclear level densities based on single-particle states that were
generated by the density-dependent Gogny interaction in the self-consistent
Hartree-Fock-Bogoliubov model. The approach of Ref. \cite{Hilaire:2001:CNL} and
our work present some analogies, which will be discussed here and in the
following sections.

The combinatorial method used in the present work is based on the Gaussian polynomial
expansion of a generating function of states of $n$ identical particles coupled to
a given angular momentum, $I$. If the particles are either bosons of the same angular
momentum, $l$, or fermions in the same $j$-shell, the problem of determining 
the multiplicities of states of given angular momentum projection, $I_z=M$, was solved long ago,
and standard solutions can be found in textbooks (see, e. g., Ref.\cite{De-Shalit:1963:NST}).

The generating function appropriate to a system of fermions in the same $j$-shell is the
following two-variable polynomial~\cite{Sunko:1985:AA}:
\begin{equation}
F(q,t)=\prod_{i=1}^{2j+1}(1+q^it)=\sum_{r=0}^{2j+1}q^{r(r+1)/2}
\left[
\begin{array}{c}
2j+1 \\
r
\end{array}
\right]_q
t^r, 
\label{eq:fgf}
\end{equation}
where the symbol in square brackets denotes denotes the Gaussian polynomial in $q$ 
of order $r(2j+1-r)$, which reduces to an ordinary binomial coefficient when $q=1$.
It can be proved that the multiplicity, $m(M)$, of states of $n$ fermions with total angular 
momentum projection $M$ is nothing but the coefficient of $q^{M+n(j+1)}t^n$ in 
formula (\ref{eq:fgf}), or, equivalently, the
coefficient of $q^M$ in 
\begin{equation}
G(2j+1,n;q)=q^{-n(2j+1-n)/2}\left[ 
\begin{array}{c}
2j+1 \\ 
n
\end{array}
\right] _q\, .  
\label{eq:gauss_poly}
\end{equation}
Finally, the number of states of total angular 
momentum $I$ is simply $N(I)=m(M=I)-m(M=I+1)$.

For the sake of completeness, we quote from Ref.\cite{Sunko:1985:AA} the generating 
function for states of bosons of angular momentum $l$:
\begin{equation}
B(q,t)=\prod_{i=1}^{2l+1}\frac{1}{1-q^it}=\sum_{r=0}^\infty q^r
\left[
\begin{array}{c}
2l+r \\
r
\end{array}
\right]_q
t^r. 
\nonumber
\end{equation}
In group-theoretical language, the above procedure corresponds to
determining the multiplicities of $SU(2)$ representations occurring in the decomposition
\[
SU(2j+1)\supset SU(2)\supset SO(2), 
\]
of the fully antisymmetric representation $\{1^n\}$ or the fully 
symmetric representation $\{n\}$ of $SU(2j+1)$ for fermions or bosons,
respectively.
The generating function approach is extended to the full Racah decomposition
\[
SU(2j+1)\supset R(2j+1)\supset SU(2)\supset SO(2), 
\]
where $R(2j+1)$ is the seniority group (orthogonal for bosons and symplectic
for fermions), on the basis of a simple consideration:
if a state of an $n$-particle configuration of given angular momentum
contains at least one zero-coupled pair, then the state appears also
in the $(n-2)$-particle configuration, because the zero-coupled pair
does not contribute to the angular momentum. As a consequence, the generating 
functions become~\cite{Sunko:1987:AA}
\begin{equation}
V_b(2l+n,n;q)=G(2l+n,n;q)-G(2l+n-2,n-2;q), 
\label{eq:bos-gen-fun}
\end{equation}
and
\begin{equation}
V_f(2j+1,n;q)=G(2j+1,n;q)-G(2j+1,n-2;q)\, ,  
\label{eq:ferm-gen-fun}
\end{equation}
for bosons and fermions, respectively, with $G(r,n;q)$ given by formula (\ref{eq:gauss_poly}). 
and $n\le(2j+1)/2$ in the fermion case, because of Pauli blocking.
The generalization of the method
to the multilevel case is straightforward: the generating function is 
the product of the generating functions (\ref{eq:ferm-gen-fun}) of individual
levels.

 This completes the formalism needed for the effective generation of
the full state space that diagonalizes the schematic Hamiltonian 
\begin{equation}
\hat{H}=\sum_i\hat{n}_i\epsilon _i+G\sum_i\hat{S}_{+}^{(i)}\hat{S}_{-}^{(i)}\, .
\label{eq:ham}
\end{equation}
Here, $\hat{n}_i$ denotes the number operator in the $i$-th single
particle level with energy $\epsilon _i$ and angular momentum $j_i$. 
$\hat{S}_{\pm }^{(i)}$ are quasi-spin operators~\cite{Ring:1980:NMB}. 
The model Hamiltonian consist
of the usual mean-field part and a diagonal pairing interaction with
constant strength $G$. The contribution of the pairing correlations
to the total energy reads 
\begin{equation}
E_P=-\frac G4\sum_i(n_i-s_i)(2j_i+1-n_i-s_i+2)\, .  
\label{eq:epair}
\end{equation}
It is important to note that our simplified residual pairing interaction 
conserves seniority, which is not realistic, but the implementation 
of the computational algorithm is straightforward.
The present method represents an extension of the model
with diagonal pairing in deformed doubly degenerate states, previously
employed by Williams~\cite{Williams:1971:AA}. 
It has been shown~\cite{Sunko:1990:AA} that the
correlations in Eq. (\ref{eq:epair}) can destabilize the shell model
ground-state, inducing a transition to a superfluid phase.

In the Gaussian polynomial method (GPM) 
the calculation is scaled down from the number of states to the
number of levels, and we emphasize that  
the method gives the exact state space for the model defined by Eq. 
(\ref{eq:ham}), taking into account the full Racah decomposition of the
many-body states with angular momentum projection $M$, seniority, energy,
and parity as good quantum numbers.

The GPM is implemented in the computer program package 
SPINDIS~\cite{Sunko:1997:AA}, which
contains various applications for calculating the total
level (or state) densities and spin-parity distributions for
the model defined by Eq. (\ref{eq:ham}). By using 
recursion relations~\cite{Sunko:1985:AA} for the
coefficients in Eq. (\ref{eq:ferm-gen-fun}),
SPINDIS allows high-speed computation. This part
of the program, called distribution generator, is served by a many-body
level-configuration generator consisting of four odometers, which control
the following quantities:

\begin{enumerate}
\item  pairs coupled to zero angular momentum, promoted among the available
levels (with Pauli blocking taken into account);

\item  unpaired particle configurations (input for the distribution
generator);

\item  number of unpaired protons (neutrons);

\item  total number of broken pairs.
\end{enumerate}

\begin{figure}[b]
 \includegraphics[angle=90,width=0.9\textwidth]{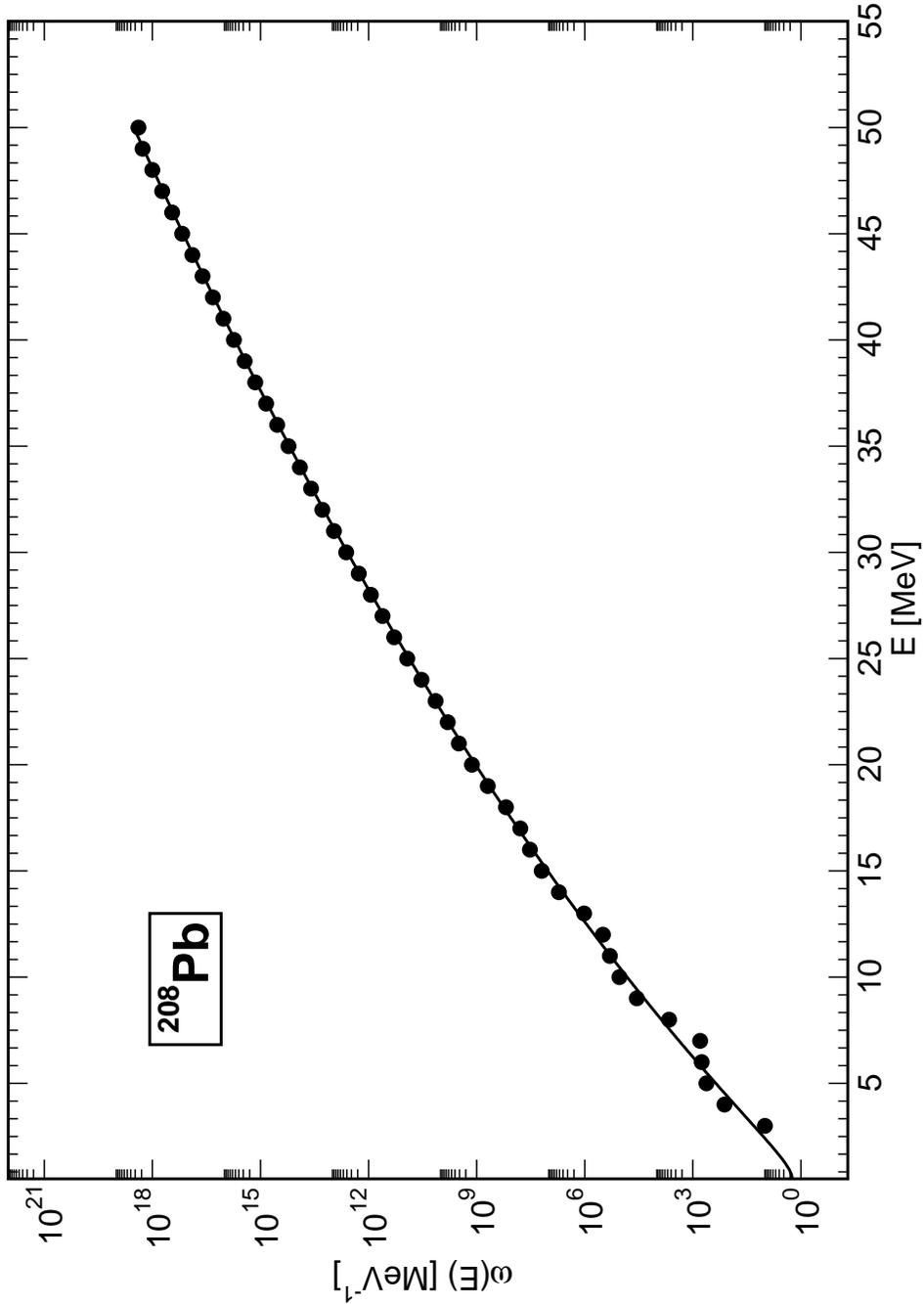}
\caption{
         Comparison of the total state density of $^{208}$Pb computed with
         the SPINDIS algorithm (dots), and with the 
	 saddle-point method (solid line),
	 using a set of realistic single particle levels
         calculated with the relativistic mean-field model with 
	 energy dependent effective mass. 
	}
\label{fig:Pb208-saddle}
\end{figure}

The energy corresponding to each many-body configuration is calculated 
by using the expression Eq. (\ref{eq:epair}) for the pairing contribution.
As it is well known, the basic
problem in the generation of multi-particle configurations is that only a
small fraction of the lexicographically generated levels is found 
below the chosen excitation energy cut-off. 
A detailed description of the algorithm for generating the
multi-particle configurations is given in Ref.~\cite{Sunko:1997:AA}. 
However, it is important to emphasize that the  
algorithm generates all possible states, i.e.  
not even a single state is lost up to the excitation
energy cut-off. This is shown in Fig. \ref{fig:Pb208-saddle},
where we compare the total state density of $^{208}$Pb up to 
50 MeV calculated by the SPINDIS algorithm with the one calculated by the
saddle-point approximation to the inverse Laplace transform of a grand-canonical 
fermion partition function.
 The latter approach to non-collective state density has been used for years,
and is known to be very reliable, provided the excitation energy is not too low. Above 
the neutron binding energy, the numerical results are in very good agreement, so that the 
two methods are mutually corroborated. The input parameters of the calculation will be 
discussed in
Section \ref{sec:num-res}. We note that at the 
excitation energy of $\approx 50$ MeV the state density 
$\omega \simeq \mathrm{\,e^{46}\,{MeV^{-1}}}$.

In addition to the total state density $\omega \left( E\right)$, SPINDIS
computes the density of states of given spin projection $\omega \left(
E,M\right)$, and the density of levels of given spin $J$
$$
\rho \left( E,J\right) =\omega \left( E,M=J\right) -
                        \omega \left( E,M=J+1\right) 
$$

What is the maximum excitation energy at which calculations can be
performed in our model? The single-nucleon wave functions are 
expanded in a spherical harmonic oscillator basis. This means that
if too many oscillator shells were included in the calculation, 
the resulting nuclear state densities would steadily increase
with energy.
 Nuclei have finite binding energies and state densities are
expected to reach a maximum well below the dissociation region, and to
decrease at higher excitation energies up to the dissociation point.
Therefore, we can either truncate the sequence of single-particle 
levels in such a way that
yields a realistic behavior of total state densities when the excitation
energy approaches the dissociation point, or we can limit our calculations to
excitation energies that are well below the expected maximum, so that the
resulting state densities are not very sensitive to the maximum number of
single-particle orbitals included in the calculations.

In the present work we have chosen the simpler alternative. The calculations
extend up to $E_{max}=50$ MeV, which is of the order of magnitude of
the nuclear potential well, and therefore of the maximum allowed one-hole
energy. This excitation energy is much lower than $E^{*}\simeq 2A$ MeV,
($A$ is the mass number)
at which the physical state density already shows significant deviations
from a simple Bethe formula like behavior.
At even higher energies, the state density
can be approximated by a Gaussian curve decreasing to 0 around 
$E^{*}\simeq 8A$ MeV~\cite{Grimes:1990:AA}. 
An application of the state density formalisms to
heavy ion reactions, where excitation energies between $2A$ and $8A$ MeV are
easily attained, necessitates a physically sound prescription for the truncation
of the single-particle level scheme. One possibility, suggested in 
Ref.~\cite{Mustafa:1992:AA}, where use is made of the recursive algorithms of
Ref.~\cite{Jacquemin:1986:AA},
 is to include, in addition to bound states, 
only those neutron quasi-bound
states that are below the centripetal barrier and the proton quasi-bound
states that are below the Coulomb plus centripetal barrier or, alternatively, 
quasi-bound states that have optical-model transmission coefficients
less than 0.05. A main default of the same work, however, is the use of a
static mean field approximation, hardly justifiable at excitation energies
above the pion production threshold, with the net result of overestimating 
the total state density at high excitation energy.

The extension of the present formalism to the energy
region of interest in heavy ion reactions will be investigated in a future
work.

\section{Collective Enhancement}
\label{sec:coll-eff}

Until now we have explicitly considered only non-collective degrees of 
freedom in generating nuclear states at high excitation energy. The
effective particle-vibration coupling that leads to realistic
single-particle spectra around the Fermi energy is treated
phenomenologically by an energy dependence of the effective 
single-nucleon potential and,
therefore, of the effective nucleon mass.

It is well known, however, that collective nuclear excitations 
give a sizable contribution to the total state
density at low excitation energy, and therefore must be included in the model 
space in order to have a meaningful
comparison with experimental data. Since the present analysis is
limited to spherical nuclei, the collective enhancement of the total state
density will be mainly due to surface vibrations of various multipolarities.
The lowest-lying vibrational modes are the quadrupole, with spin-parity $L^\pi
=2^{+}$ and energy $E\left( 2^{+}\right) \simeq 30A^{-2/3}$ MeV, and
the octupole, with $L^\pi =3^{-}$ and $E\left( 3^{-}\right) \simeq 50A^{-2/3}$
MeV.

The effect of collective degrees of freedom on total state densities is
conveniently illustrated by assuming the adiabatic decoupling of
collective and non-collective excitations, approximately valid at excitation
energies $E$ much higher than the average, or characteristic, collective
energy $\left\langle E_c\right\rangle $. This assumption is generally not valid
in the energy range of interest to the present work, and is made
here only for the sake of simplicity.

Let $\left\{ E_c\right\} $ be the set of eigenvalues of a collective 
Hamiltonian. For any given eigenvalue $E_c$,
the effective excitation energy available to non-collective modes,is $%
E^{*}=E-E_c$, and the total state density is obtained by summing the
combinatorial state densities corresponding to all effective excitation
energies 
\begin{eqnarray}
\omega _{tot}\left( E\right) &=&\sum_c\omega \left( E-E_c\right)  
\nonumber \\
&\simeq &\sum_c\left[ \omega \left( E\right) -E_c\frac{\partial \omega }{%
\partial E}\left( E\right) \right] \, ,
\label{sd-tot}
\end{eqnarray}
provided that $E_c\ll E$. Let us introduce the temperature $T$
corresponding to the excitation energy $E$, according to the definition 
\begin{equation}
\frac 1T=\frac \partial {\partial E}\ln \omega \left( E\right) =\frac 1\omega 
\frac{\partial \omega }{\partial E}\left( E\right) \, .  \label{T_mc}
\end{equation}
Eq. (\ref{sd-tot}) can thus be rewritten in the form 
\begin{eqnarray}
\omega _{tot}\left( E\right) &\simeq &\omega \left( E\right) \sum_c\left( 1-%
\frac{E_c}{T\left( E\right) }\right)  
\nonumber \\
&\simeq &\omega \left( E\right) \sum_c\exp \left( -\frac{E_c}{T\left( E\right) 
}\right) 
\nonumber \\
&=&\omega \left( E\right) Z_{coll}\left( E\right) \, .
\nonumber
\end{eqnarray}
The enhancement factor is the collective partition function 
$Z_{coll}\left( E\right) $ at temperature $T\left(E\right) $.

We emphasize that the above derivation is not based on the
assumption that the fermion system is in equilibrium with an external heat
and particle bath, as required by a grand-canonical ensemble description. On
the contrary, our fermion system is isolated, since the non-collective state
density $\omega $ is calculated by a combinatorial method consistent with a
micro-canonical ensemble description, and the definition (\ref{T_mc}) of
temperature is based on Boltzmann's definition of entropy, $S\left( E\right)
=\ln \omega \left( E\right) +c$, valid in the micro-canonical ensemble.

\begin{figure}[b]
 \includegraphics[angle=90,width=0.9\textwidth]{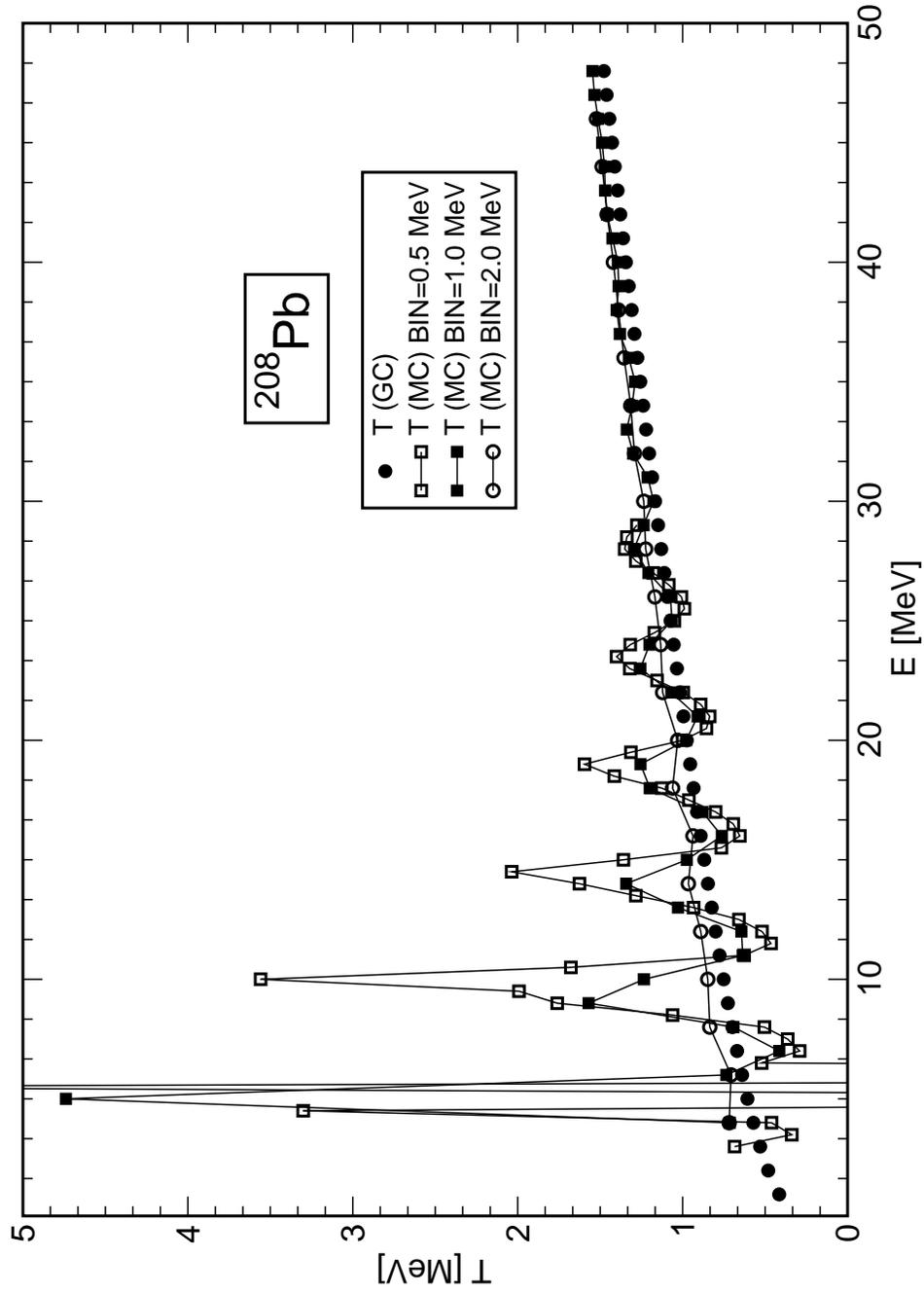}
\caption{
         Comparison of the energy averaged micro-canonical temperature $T$
	 (for three different values of the energy bin), and the 
         grand-canonical temperature $T_{gc}$, as functions 
	 of $E$ for $^{208}$Pb.
         }
\label{fig:Pb208T}
\end{figure}

As was pointed out in Ref. \cite{Morrissey:1994:AA}, 
the above micro-canonical description
can be transformed into a grand-canonical one by observing that $S\left(
E\right) =S_{gc}\left( E\right) +\Delta S\left( E\right) $, where $\Delta
S\left( E\right) $ goes to zero for $E \rightarrow \infty$. Thus, 
\begin{equation}
\frac 1{T_{gc}}\equiv \frac{\partial S_{gc}}{\partial E}\left( E\right)
=\frac 1T-\frac{\partial \Delta S}{\partial E}\left( E\right) >\frac 1T \, .
\label{T-gc}
\end{equation}
As a consequence we expect that, as soon as the excitation energy is high
enough for a grand-canonical description to become applicable, 
the fermion system can
be considered in thermal equilibrium with an external reservoir at a
temperature $T_{gc}\left( E\right) $, systematically lower than the
micro-canonical temperature $T$, derived from the combinatorial state
density $\omega $, at the same energy $E$. 
This is confirmed by Fig. \ref{fig:Pb208T},
which compares the grand-canonical and micro-canonical temperatures of 
$^{208}$Pb as functions of $E$. We note that the
micro-canonical temperature has been obtained by numerical derivation of an
energy-smoothed state density, in order to avoid numerical divergence at
energies where the original combinatorial state density displays 
discontinuities connected with the opening of particle-hole excitation
channels. More precisely, formula (\ref{T_mc}) predicts  divergence of 
inverse temperature, and, consequently, a sudden temperature dip at any
discontinuity connected with a new particle-hole excitation, followed by a zero
inverse temperature, and then a diverging temperature, in the plateau between
two successive discontinuities of the state density. The rapid oscillations
of the micro-canonical temperature derived from a state density computed in
energy bins, $\Delta E$, as small as 500 KeV, are progressively washed out  
with increasing $\Delta E$ up to 2 MeV, where good agreement is reached with
the grand-canonical temperature corresponding to the same excitation energy.

A non monotonic behaviour of temperature as a function of excitation energy
was experimentally observed for the first time in the level densities of
well deformed
lanthanides below 6 MeV~\cite{Melby:1999:AA} and, more recently,
also in the level densities of weakly deformed Sm isotopes~\cite{Siem:2002:AA}
and  attributed to
progressive breaking of
nucleon pairs, and, at higher energies, quenching of pair correlations.
Although this effect has not been observed till now in magic nuclei, where the
discontinuities in level densities should be connected with non-collective
particle-hole excitations across shell closures,  a proper
extension of the micro-canonical formalism to open-shell nuclei would make
it possible to compare calculated temperature fluctuations with those
extracted from experimental data, thus giving an important advantage at low
excitation energies over a canonical formalism, where temperature 
fluctuations are smoothed out by construction.

If the derivative in Eq. (\ref{T-gc}) is approximated by 
\begin{equation} \nonumber
\frac{\partial \Delta S}{\partial E}\simeq \frac{\delta S}{\delta E} \, ,
\end{equation}
the expression can be rewritten as 
\begin{equation} \nonumber
\frac{\delta E}T\simeq \frac{\delta E}{T_{gc}}-\delta S \, , 
\end{equation}
i.e. it relates a change in excitation energy in the micro-canonical system
with a change of energy and entropy in the grand-canonical frame.

Once we are allowed to reformulate the equilibrium of our fermion system in
a grand-canonical language, it becomes natural to express the collective
enhancement factor, $Z_{coll}$, in terms of a boson system in equilibrium
with fermions at temperature $T_{gc}\left( E\right) $. The bosons represent
collective fermion pairs treated in the random phase approximation (RPA), as
shown in Ref. \cite{Ignatyuk:1975:AAb}, where the vibrational factor is 
expressed in terms of the energy and entropy of a gas of RPA 
phonons of both parities $\pi_i $, and all multipolarities $\lambda _i$ 
\begin{equation} \nonumber
Z_{coll}\left( T_{gc}\left( E\right) \right) =\exp \left( \delta S-\frac{%
\delta E}{T_{gc}}\right) \, ,  
\end{equation}
where 
\begin{eqnarray}
\delta E &=&\sum_i\sum_{\pi _i}\left( 2\lambda _i^{\left( \pi _i\right)
}+1\right) n_i \omega _i, 
\nonumber \\
\delta S &=&\sum_i\sum_{\pi _i}\left( 2\lambda _i^{\left( \pi _i\right)
}+1\right) \left[ \left( 1+n_i\right) \ln \left( 1+n_i\right) -n_i\ln
n_i\right] \, . 
\nonumber
\end{eqnarray}
$ \omega _i$ is the energy of the RPA phonon of type $i$, with
spin-parity $\lambda _i^{\left( \pi _i\right) }$. In the present analysis we
only consider the leading modes of both parities, namely the quadrupole ( $%
\lambda _i^{\left( \pi _i\right) }$ $=2^{+}$ ) and the octupole ( $\lambda
_i^{\left( \pi _i\right) }=3^{-}$ ). The average number of type-$i$ phonons
is given by a Bose distribution corrected for damping effects 
\cite{Lunev:1999:AA}: 
\begin{equation}
n_i\left( T\right) =\frac{\exp \left[ -\frac{\gamma _i\left( T\right) }{%
2 \omega _i}\right] }{\exp \left( \frac{ \omega _i}T\right) -1}\, ,
\label{eq:nb}
\end{equation}
where the temperature dependence of the damping factor  $\gamma _i$ 
is taken from the Landau theory of Fermi liquids.
The numerical coefficients are determined by a
consistent description of collective levels and neutron resonance 
data~\cite{Blokhin:1988:AAa} 
\begin{equation} \nonumber
\gamma _i\left( T\right) \simeq 0.0075A^{\frac 13}\left[ \left(  \omega
_i\right) ^2+4\pi ^2T^2\right] \hspace{0.25in}{\rm MeV}  
\label{eq:gam}
\end{equation}
In the limit ${T \rightarrow \infty }$, Eqs. (\ref{eq:nb}) and (\ref{eq:gam})
give $n_i \rightarrow 0$ and, consequently  
$\lim_{T\rightarrow \infty }$ $Z_{coll}=1$. 

In principle, because the combinatorial algorithm~\cite{Sunko:1985:AA}
exploited by SPINDIS can be also applied to bosons, 
as illustrated in the previous section,
the vibrational enhancement can be described directly in the
micro-canonical frame. A similar procedure
is followed in Ref. \cite{Hilaire:2001:CNL}, where a three-variable generating
function is used, whose expansion coefficients give the 
multiplicities of states of
given boson number and spin projection at a given excitation energy. In the
latter approach, however, the damping of collectivity , which is quite natural
in the finite temperature description of Eq. (\ref{eq:nb}), is introduced
by truncating \textit{ad hoc} the boson generating function at $n_{max}=3$,
with the justification that states with higher boson number are rather weakly
collective.

Neither of the two approaches to the vibrational enhancement is, however,
very satisfactory from a theoretical point of view.
Both treat the excited nucleus as an assembly
of independent fermions and bosons. This approximation might be acceptable,
and is confirmed \textit{a posteriori } by comparison with experimental data,
if we limit ourselves to the calculation of total
level densities. 
A reliable description of partial level densities of given spin and parity,
on the other hand, necessitates a consistent treatment of the boson-fermion
interaction, by taking properly into account the microscopic structure of
bosons as collective fermion pairs.

The present analysis is limited only to spherical nuclei, and, consequently,
only the vibrational enhancement of state densities is taken into account,
because our main goal is to check the validity of the relativistic
mean-field approach to excited systems where correlation effects, not yet
easily treatable in a relativistic framework, play a minor role.
As is known, 
the extension to deformed nuclei will require, in addition, a rotational
enhancement, as well as an appropriate treatment of the
coupling of rotations and vibrations.
The latter coupling is frequently neglected in level density calculations,
thus leading to a significant overestimate of the collective enhancement
around its maximum, as shown in a recent evaluation~\cite{Mengoni:2002:AA} 
of this effect in the framework of the interacting boson model.

A fully grand-canonical description of 
both  single-particle and collective degrees
of freedom in nuclear level densities smooths out the low energy fluctuations
connected with the onset of non-collective particle-hole 
excitations and approaches the
micro-canonical description with increasing excitation energy.

      A self-consistent mean-field approach to nuclear level densities in the
grand-canonical framework is presented in 
Ref.~\cite{Demetriou:2001:AA}. 
There, the single-nucleon states are calculated in the Skyrme-Hartree-Fock
approximation by means of the MSk7 effective interaction
 and the pairing interaction is treated in the
standard BCS approximation. The average effective mass turns out to be
$<m^* \simeq 1.05 m$.
A temperature-dependent rotational enhancement of standard form is taken into
account in the level densities of strongly deformed nuclei.
Although the calculation 
does not include any explicit vibrational enhancement for spherical nuclei, 
this effect is implicitly taken
into account by adjusting excitation energy and entropy so as to reproduce
empirical data, such as the cumulative numbers of discrete levels and the 
neutron resonance spacings. This study, however, is neither limited to the about
300 nuclei whose neutron resonance spacings are measured, nor
to the about 1200 nuclei whose discrete spectra are known to some extent, but,
aiming at astrophysical applications, is  
extended to 8000 nuclei outside the beta-stability valley, where extrapolation
of empirical systematics would not make any sense. 
Total and spin-dependent level
densities are provided in tabular form up to an excitation energy of 150 MeV.

      A new promising approach to nuclear level densities in the
canonical framework is provided by the shell-model Monte Carlo      
(SMMC) method~\cite{Nakada:1997:AA}. There, use is made of the 
Hubbard-Stratonovich representation for the $e^{-\beta H}$ operator, $\beta$ 
being the inverse temperature, and the energy of the system , $E$, is computed
by the auxiliary field method
as a function of $\beta$ from the canonical expectation value of the 
Hamiltonian through an exact particle-number projection of both protons and
neutrons. The partition function, $Z(\beta)$, is then determined by numerical
integration of $E(\beta)$ and the level density obtained as the inverse
Laplace transform of $Z$, in the saddle-point approximation.
      Using in their Hamiltonian a separable 
surface-peaked residual interaction, in 
addition to the standard pairing interaction, active in the $(pf+0g_{9/2})$
valence shell, the authors of Ref.~\cite{Nakada:1997:AA} were able to 
reproduce the experimental level density of ${}^{56}$Fe up to 20 MeV, thus 
treating collective and non-collective excitations on the same computational
basis, without resorting to any external enhancement factor. 
The parity-projected level densities in the above mentioned range 
 were well reproduced by  back-shifted Bethe formulae with parity-dependent
parameters.  
     
    The SMMC method was later applied to other even-even nuclei, from Fe to
Ge~\cite{Nakada:1998:AA}, as well as to odd-mass and odd-odd 
nuclei~\cite{Langanke:1998:AA} in the same mass region and energy range.
In the case of odd fermion systems, the weight functions appearing in the
auxiliary field integrals are not necessarily positive, thus giving rise to
a well-known sign problem, recently overcome by the particle-number
reprojection method~\cite{Alhassid:1999:AA}.
    All the above mentioned SMMC calculations are well reproduced
 by back-shifted Bethe formulae, which are thus able to fit theoretical
 level densities which
include correlations among nucleons, at least for nuclei which are not strongly
deformed, in an energy range of the order of some ten MeV.

     The possibility to include correlations in the Bethe formula has been
known empirically for years, and is reflected in the fact, already quoted in
Section 2, that the average value of the level density parameter at the
neutron binding energy, $a = A/8$ MeV${}^{-1}$, is about as twice as the value
predicted by the Fermi gas model, thus accounting for low energy correlations
other than those described by the energy backshift.
 A theoretical justification
of the validity of the shifted Bethe formula beyond the independent particle
approximation was recently given in Ref.~\cite{Zuker:2001:AA}. There, it was
shown that the shifted Bethe formula is closely connected with a continuous
binomial level density in a finite space, defined by three parameters, which
can be extracted from the three lowest moments of a Hamiltonian of any rank.
     
We believe, however, that the mean-field approach to nuclear level densities
is still worthy of development and applications, were it not for the fact that
the SMMC method is still very far from applicability to a region of the
periodic chart and to an energy range as large as those 
allowed in a self-consistent mean-field approximation (see, in particular,
Ref.~\cite{Demetriou:2001:AA}). Moreover, at excitation energies higher than
about 2 MeV/nucleon, easily attainable in current heavy ion reactions, we 
expect not only vanishing correlations, but also failure of the Bethe formula,
which predicts an exponentially increasing level density, physically 
unreasonable at excitation energies approaching the nuclear binding energy, as
already discussed at the end of the previous section. Under such extreme
conditions, microcanonical, or canonical calculations based on a self-consistent
relativistic mean-field
approximation may prove a very valuable approach, as we intend to show in a
future work.

\section{Numerical Results}
\label{sec:num-res}

In this section we discuss and compare with experimental data
the results of illustrative calculations of total level densities of 
few magic and semi-magic nuclei.

A crucial parameter of our calculations is the density of 
single-particle states at the Fermi energy. The single-nucleon 
spectra are calculated with the relativistic mean-field model with
energy-dependent effective mass. For the initial 
parameterization of the effective Lagrangian (meson masses and 
meson-nucleon couplings) we have chosen, 
instead of the NL3 effective interaction~\cite{Lalazissis:1997:AA} 
adopted in Ref.~\cite{Vretenar:2002:PDN}, the older parameter set NL1
\cite{Reinhard:1986:AA}. Even though NL1 is admittedly
inferior to NL3 in reproducing the bulk nuclear properties,
it yields a more realistic
magic gap $E\left( h_{9/2}\right) -E\left( s_{1/2}\right) $ in $^{208}$Pb,
considered also in the present work. For this reason, 
NL1 has also been adopted as
the starting parameterization in the recent analysis \cite{Niksic:2002:AA}
of shape coexistence phenomena in the Pt-Hg-Pb region.

In addition, the half-width of the interval in which the 
$\sigma $ and $\omega $ mean-field potentials 
are assumed to be linearly dependent on energy, has been reduced from
10 to 5 MeV. In this way, we obtain more realistic energies of the
single-particle states above the Fermi surface,
which have a particularly
strong effect on the calculated total level densities.

The linear energy dependence of the $\sigma $ 
and $\omega $ mean-field potentials is determined by the parameter $\alpha$  
in Eq. (\ref{eq:VE}).

The input parameters used in the calculation of three illustrative
cases, ${}^{208}$Pb, ${}^{114}$Sn and ${}^{60}$Ni, are collected in 
Table \ref{tab:mf-par} (parameters of the relativistic mean-model), and 
Table \ref{tab:calc-par} (the parameters $\alpha$ of the energy-dependent
effective mass, the pairing strengths and the phonon energies). 

Starting from the heaviest nucleus, $^{208}$Pb, $\alpha$ has been adjusted
so as to reproduce 
the binding energy, the cumulative
number of discrete levels as a function of excitation energy, and the $s-$wave 
neutron resonance spacing, $D_{obs}$, measured
in the $^{207}Pb+n$ reaction. As is known,
in the case of an odd-mass target nucleus with $N$ neutrons and ground-state
spin-parity $I^{\pi}$,
\begin{equation}
D_{obs} =\frac{1}{\rho(B_n,J=I+1/2,\pi)+\rho(B_n,J=I-1/2,\pi)},
\label{D_obs}
\end{equation}
where $B_n$ and $\rho(U,J,\pi)$ are, respectively,
the neutron binding energy and the density of levels with spin-parity
$J^{\pi}$ of the compound nucleus with $N+1$ neutrons.  
Since no experimental data are available on the total level density 
at higher excitation energy, we have checked the quality of our
calculations by comparing the total state density obtained from the
SPINDIS algorithm applied to the set of experimental single-particle
levels, already adopted in Ref.~\cite{Vretenar:2002:PDN}, with the
analogous level density obtained with the same algorithm applied to
the single-particle levels generated in the RMF approximation with
non-zero $\alpha$ parameter. The agreement is excellent over the whole
energy range considered in the present work.

The theoretical non-collective level density thus obtained has been multiplied
by the vibrational enhancement factor described in the previous section, by
assuming quadrupole and octupole phonon energies close to the energies
of the $2^+_1$ and $3^-_1$ levels, respectively, somewhat different, for
a bi-magic nucleus, from the values expected from systematics.
 
\begin{figure}[b]
 \includegraphics[angle=90,width=0.8\textwidth]{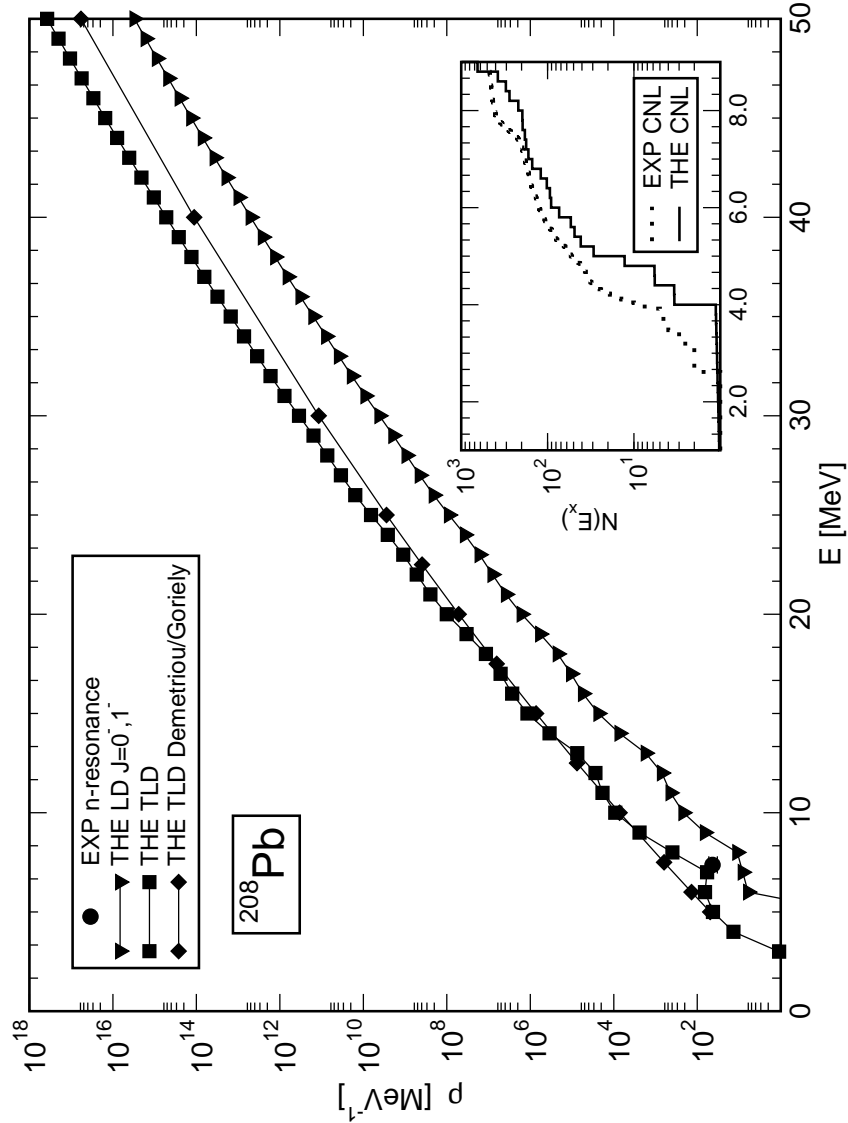}
\caption{
         Level densities of $^{208}$Pb as functions of the 
	 excitation energy: a) total level
         density (TLD) (squares); b) sum of the partial level densities for
	 $J^\pi =0^{-}$ and $J^\pi =1^{-}$ (triangles), compared 
	 with the experimental $s$-wave
	 neutron resonance density of the $^{207}Pb+n$ reaction 
	 (circle)~\protect\cite{Mengoni:1994:AA};
         c) total level density from Ref.~\cite{Demetriou:2001:AA}
         (diamonds). In the inset 
	 the calculated cumulative numbers of discrete levels (CNL)
	 (solid line) are shown in comparison with 
	 the experimental values~\protect\cite{ENSDF:2002:OL}.
         }
\label{fig:Pb208}
\end{figure}

The resulting total state density is compared in Fig. \ref{fig:Pb208}
with the corresponding grand-canonical calculation
 of Ref.~\cite{Demetriou:2001:AA}. The agreement appears to be
satisfactory up to about 20 MeV, while our level density becomes larger
than that of Ref.~\cite{Demetriou:2001:AA} with increasing excitation energy.
It is worthwhile to stress, however, that Ref.~\cite{Demetriou:2001:AA} does not
include an explicit vibrational enhancement factor for spherical nuclei.
 Experimental and calculated cumulative numbers of discrete levels
at low excitation energy are compared in the inset of the same figure.

In order to check the possibility of reproducing our numerical results with
simple analytic formulae, we have compared the energy trend of our total
state density with
the standard backshifted Bethe formula (BBF)
\cite{Egidy:1986:AA}:
\begin{equation} \label{eq:bbf}
\omega ^{\rm{BBF}} _{\rm{T}} (E)= g \frac{\sqrt{\pi}}{24}
\frac{\exp \left[{2 \sqrt{a(E-E_0)}} \right]}{
               a^{1/4} (E-E_0)^{5/4}} \, ,
\end{equation}
and with the generalized Bethe formula (GBF) \cite{Paar:1997:AAb}:
\begin{equation} \label{eq:gbf}
\omega ^{\rm{GBF}} _{\rm{T}} (E)=
        g\frac{\sqrt{ \pi (1-\xi) }}{6 \sqrt{8}}
         \cdot \frac{\exp \big\{[a(E-E_0)]^{\xi}/{\xi} \big\}}
         { (E-E_0)[a(E-E_0)]^{1 - 3 \xi /2}} \, .
\end{equation}
In both formulae, the total state density is obtained for $g=2$. Moreover,
GBF reduces to BBF for $\xi=0.5$.

\begin{figure}[b]
 \includegraphics[angle=90,width=0.9\textwidth]{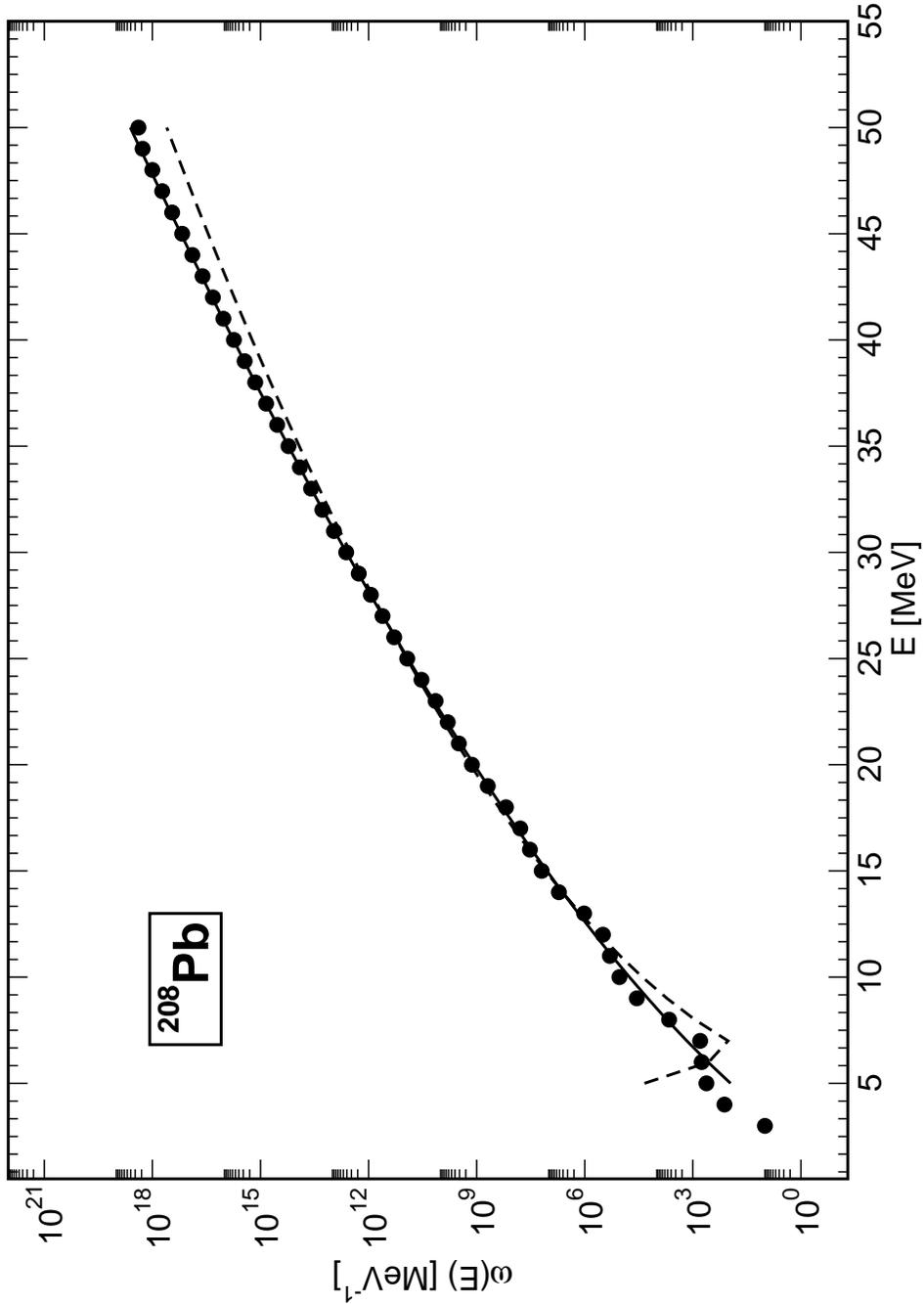}
\caption{
         Total state density of $^{208}$Pb computed with
         the SPINDIS algorithm (dots)
         and fitted by BBF (\protect\ref{eq:bbf})
	 (dotted line) and GBF (\protect\ref{eq:gbf}) (solid line) in
	 in the [5,30] MeV interval. The BBF and GBF curves are
	 extrapolated to the higher energy region for the sake of
	 comparison with SPINDIS.
	}
\label{fig:Pb208-fit}
\end{figure}

The BBF and GBF fits to the numerical state density of $^{208}$Pb are
shown in Fig. \ref{fig:Pb208-fit} and the corresponding parameters,
adjusted on the energy range from 10 to 30 MeV, are given
in Table \ref{tab:fit-par}. Our numerical results are well reproduced
by GBF on the whole energy range up to 50 MeV, while BBF exhibits sizable
discrepancies outside the region of fit. The BBF curve appears to be in
better agreement with the 
total state density of Ref.~\cite{Demetriou:2001:AA}, 
probably in connection with the fact that the latter does
not include an explicit vibrational enhancement.

\begin{figure}[b]
 \includegraphics[angle=90,width=0.9\textwidth]{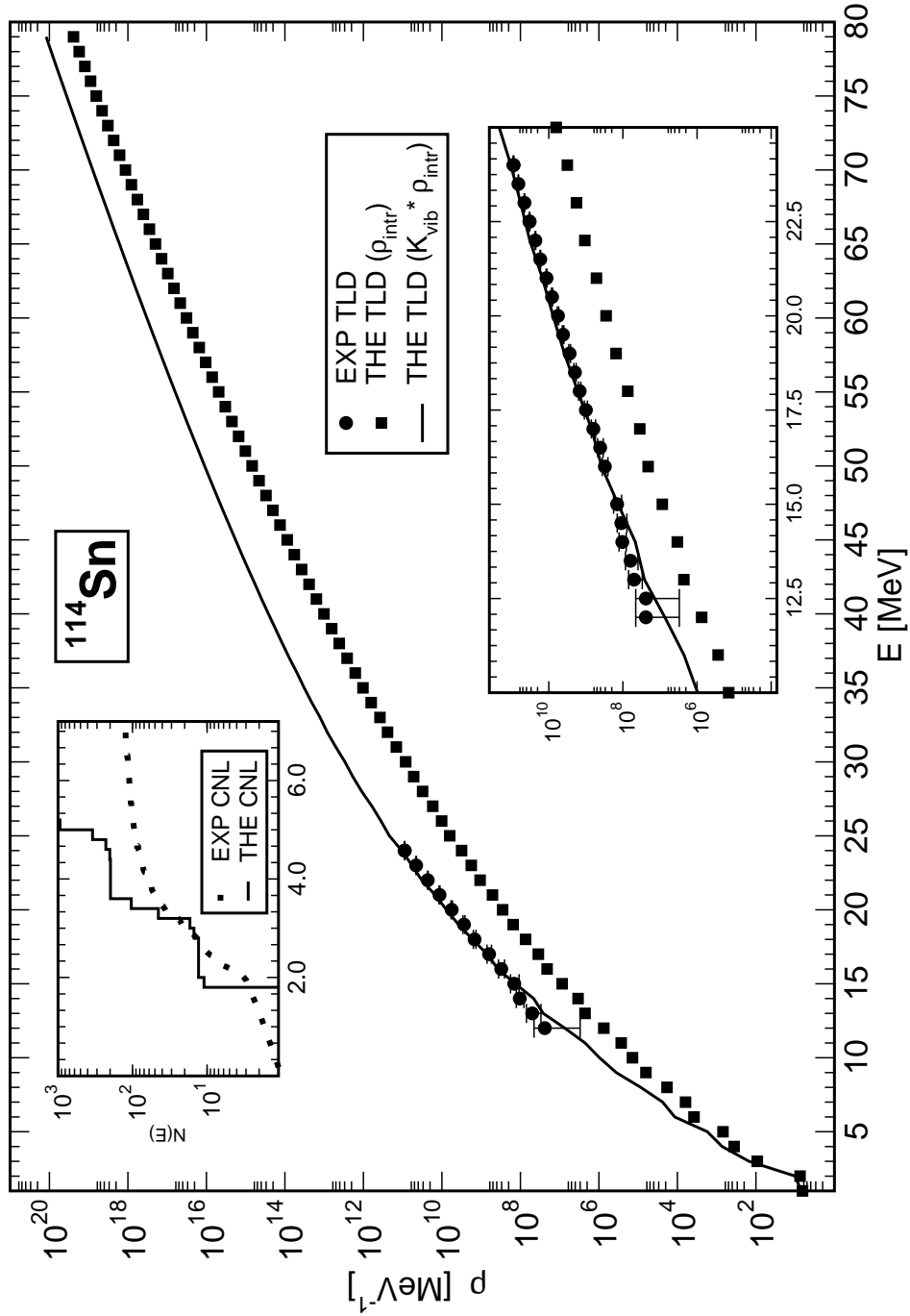}
\caption{
         Total level density (TLD) of $^{114}$Sn as function of the 
	 excitation energy, calculated with (solid line) 
	 and without the vibrational enhancement factor (squares), and compared 
	 with experimental data from Ref. \protect\cite{Chakrabarty:1995:EVN}.
	 Lower right inset: the same results magnified with all
	 available experimental points present.
	 In the upper inset 
	 the calculated cumulative numbers of discrete levels (CNL)
	 (solid line) are shown in comparison with 
	 the experimental values~\protect\cite{ENSDF:2002:OL}.
         }
\label{fig:Sn114}
\end{figure}

\begin{figure}[b]
 \includegraphics[angle=90,width=0.9\textwidth]{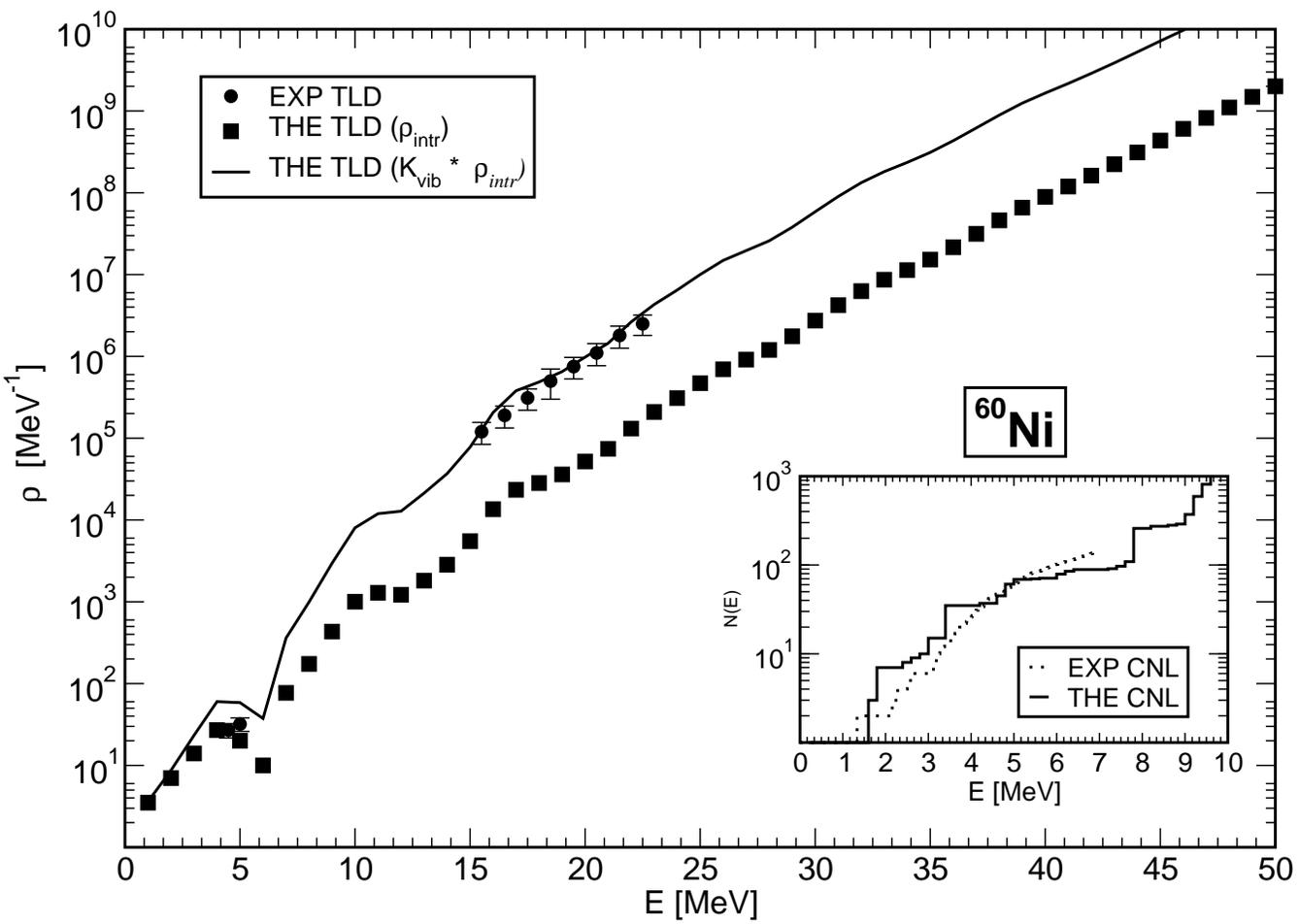}
\caption{
	 The same as in Fig. \protect\ref{fig:Sn114}, but for $^{60}$Ni.
	 The experimental data for the level density are from  
         Ref. \protect\cite{Iljinov:1992:PSA}. 
	          }
\label{fig:Ni60}
\end{figure}

For lighter spherical nuclei, there are experimental data on total level
densities mainly below 20 MeV, where the
vibrational enhancement factor discussed in 
Section \ref{sec:coll-eff} is expected to play a
significant role. This is indeed the case for the semi-magic nuclei $^{114}$Sn
and $^{60}$Ni, whose calculated level densities are compared with
experimental data in Figs. \ref{fig:Sn114} and \ref{fig:Ni60}, respectively. 
Here, $\alpha$ is adjusted again on the binding energy and the cumulative
number of discrete levels, while the level density at higher energy is
reproduced by means of the smooth vibrational factor discussed in the previous
section.

The total state densities including collective effects are equally well
reproduced by BBF and GBF with the parameters given in Table 
\ref{tab:fit-par} up to about 40 MeV, while the damping of the collective
factor with increasing energy yields systematically lower results than BBF
and GBF at higher excitation. These results are in qualitative agreement with
the conclusions of Ref.~\cite{Zuker:2001:AA} about the range of validity of
BBF.

The relativistic mean-field model, the combinatorial algorithm and the
vibrational enhancement factor adopted in the present work are strictly
valid for spherical nuclei only. In order to investigate the possible
extension to open shell nuclei, we have performed calculations also for
slightly deformed nuclei in the $A = 60$ mass region, in particular
${}^{56}$Fe and ${}^{55}$Mn, for which experimental data of the same quality
as ${}^{60}$Ni are available. 

\begin{figure}[b]
 \includegraphics[angle=90,width=0.9\textwidth]{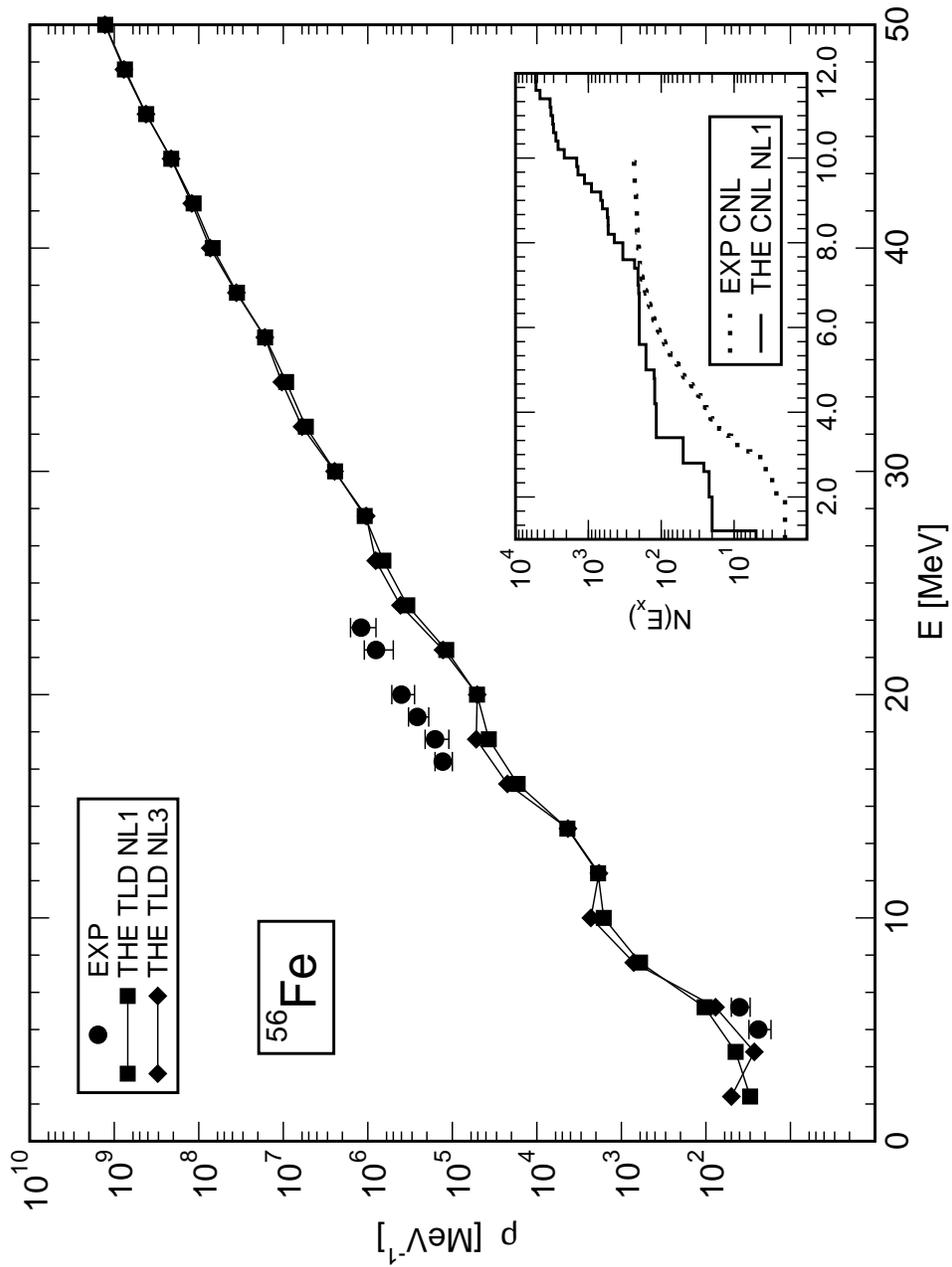}
\caption{
	 Level densities of $^{56}$Fe as functions of the 
	 excitation energy: 
	 a) The experimental data for the level density are from  
         Ref. \protect\cite{Lu:1972:LDN,Iljinov:1992:PSA}
	 (circles). 
	 b) total level density (TLD) calculated using NL1 
	 parametrisation (squares); 
	 c) total level density (TLD) calculated using NL3 
	 parametrisation (diamonds);
	 In the inset the calculated cumulative numbers of 
	 discrete levels (CNL) (solid line) are shown in 
	 comparison with the experimental 
	 values~\protect\cite{ENSDF:2002:OL}.
	          }
\label{fig:Fe56}
\end{figure}

\begin{figure}[b]
 \includegraphics[angle=90,width=0.9\textwidth]{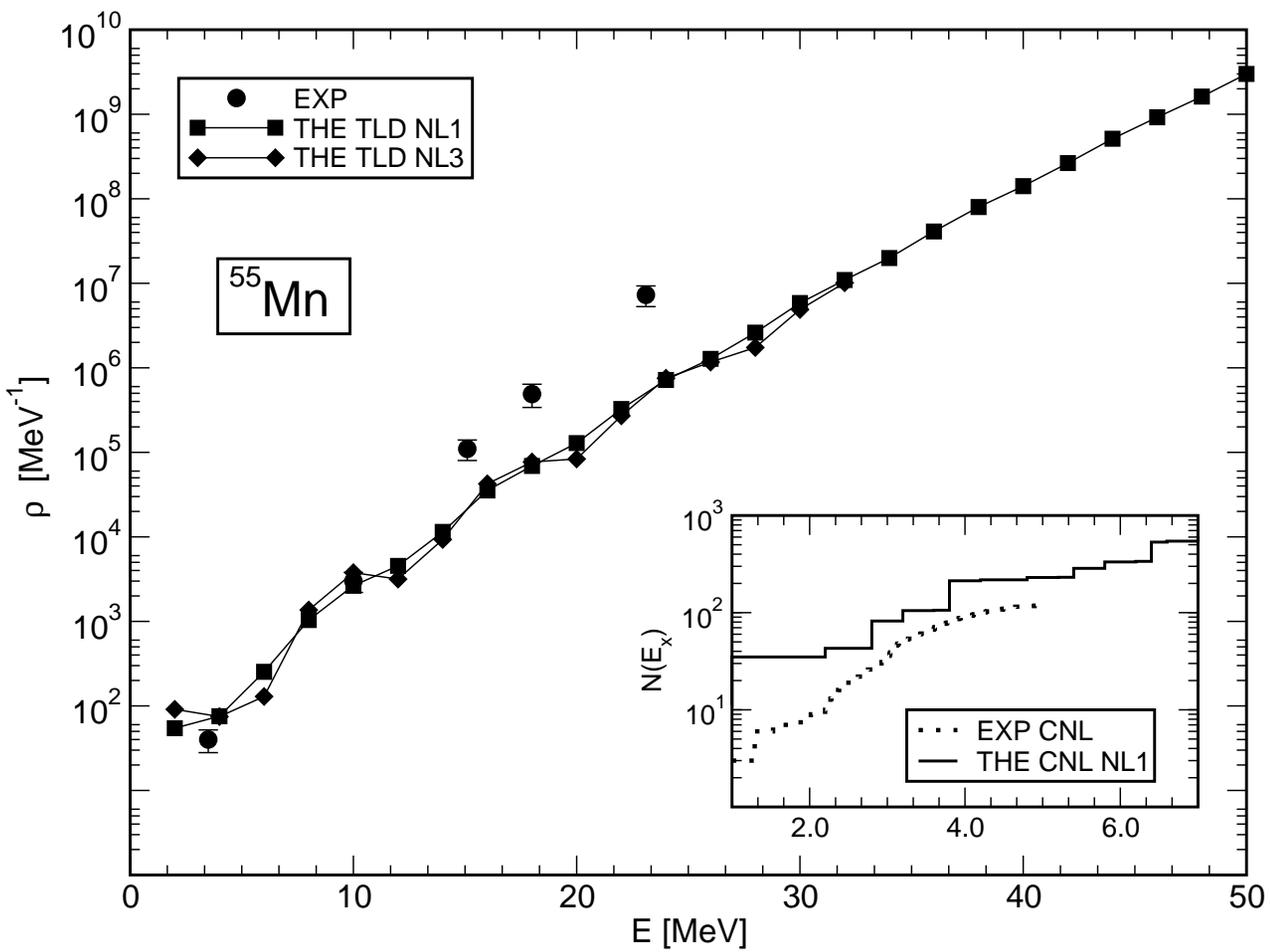}
\caption{
	 The same as in Fig. \protect\ref{fig:Fe56}, but for $^{55}$Mn.
	          }
\label{fig:Mn55}
\end{figure}

Comparison of calculated and experimental level densities, shown in Figs.
\ref{fig:Fe56} and \ref{fig:Mn55} for ${}^{56}$Fe and ${}^{55}$Mn,
respectively, indicates, however, that for open-shell nuclei there are 
some problems still to be solved. To bring these problems into focus, 
level densities have been calculated with $\alpha$ = 0 and are plotted
without any collective enhancement factor.

In general, total level densities below 10 MeV turn out to be slightly overestimated
by our calculation, as a consequence of the degeneracy of the ground state,
as shown by the insets of Figs. \ref{fig:Fe56} and 
\ref{fig:Mn55}, where calculated and experimental
cumulative number of levels are compared.  
The low energy data are derived from the cumulative number of discrete 
levels, whose spectra are somewhat different from those of spherical 
nuclei. In particular, both ${}^{56}$Fe and the even-even core of 
${}^{55}$Mn, i. e. ${}^{54}$Cr, show a ground-state rotational band 
including the $2^+_1$, $4^+_1$, $6^+_1$ and $8^+_1$ states, while the
energy of the  $2^+_2$ state is of the order of magnitude of the
quadrupole phonon energy from phenomenological systematics. 

It is not expected that the single-particle level scheme from a
mean field with spherical symmetry and the combinatorial SPINDIS
algorithm for spherical nuclei adopted in the present work produce a
very realistic distribution of non-collective states at low energy, neither 
can a simple vibrational enhancement factor compare well with the 
experimental level densities above 15 MeV, where both vibrational
and rotational degrees of freedom are simultaneously excited and
coupled together.

Therefore, necessary conditions for extension of the present mean-field 
approach to open-shell nuclei appear to be the evaluation of single-particle 
level schemes at realistic deformations and the use of a combinatorial 
algorithm, such as the recursive one~\cite{Williams:1969:AA,Jacquemin:1986:AA},
valid also for deformed systems. 
Once a more realistic single-particle scheme is obtained, more reliable
conclusions can be drawn about the treatment of the particle-vibration coupling,
simulated by the $\alpha$ parameter of our model. Corresponding
calculations~\cite{Donati:2000:A}
of the effective nucleon mass in a cranked Nilsson-BCS approach to the
structure of statically deformed nuclei show that the frequency dependent 
factor, the so-called $\omega$ mass, $m_{\omega}$, of the effective nucleon mass 
is somewhat different from that obtained for open-shell nuclei with
spherical symmetry, i. e. the one that could be simulated with
the present version of our model.

 Moreover, the derivation of a collective enhancement factor
for deformed nuclei is nontrivial and requires
an investigation of the effect of coupling of different
collective modes, such as rotations and surface vibrations,
not only with the fermion degrees of freedom, but also among themselves.
The latter coupling, often neglected in the level density formalisms, is 
expected to play an important role in the energy range considered in the 
present work. 

For the sake of comparison, we show in Figs. \ref{fig:Fe56} and \ref{fig:Mn55}
also the level density resulting from the NL3 parametrization given in
Tab. \ref{tab:mf-par}. NL1 turns out to be in slightly better overall 
agreement with the experimental data.

\section{Conclusions and Perspectives}
\label{sec:concl}

The relativistic mean-field model with energy-dependent effective mass, 
introduced in Ref. \cite{Vretenar:2002:PDN}, exhibits
realistic densities of single-nucleon states at the Fermi
energy. When coupled to the combinatorial algorithm SPINDIS for 
treating non-collective excitations of a spherical nucleus,
and to a simple vibrational partition function for
collective excitations, the model yields  
total level densities in agreement with
experimental data in magic and semi-magic nuclei. 

In the present form, however, the model contains a phenomenological
treatment of surface vibrations and of their coupling to the
single-particle degrees of freedom. A fully microscopic formulation
would require generation of phonons in a relativistic RPA
approximation and explicit formulation of the particle-vibration coupling,
presently simulated by the linear energy dependence of the meson fields.

The planned extension to non-magic deformed nuclei, necessary
for the present approach to become competitive with the
non-relativistic mean-field approaches
\cite{Hilaire:2001:CNL,Demetriou:2001:AA}, requires calculation
 of deformed single-particle
schemes, an appropriate combinatorial algorithm and a collective 
factor taking into account rotational and vibrational degrees of
freedom, and their coupling as a function of excitation
energy, or temperature.  

There is, moreover, a kind of level density problem
where the relativistic mean-field approach should have no 
true competitors, i. e. the calculation of nuclear level
densities above the threshold of pion production and up
to the total binding energy of the system, where a
static non-relativistic mean-field approximation loses its
validity. 
The present linear energy dependence of the classical meson fields
is, admittedly, a low-energy approximation, to be properly modified 
on the basis of the results of the relativistic Br\"uckner-Hartree-Fock 
theory~\cite{Jaminon:1990:AA}, which predicts a faster decrease of the
real potential well with increasing excitation energy and, consequently,
a sizable change of the single-particle level scheme, as well as the
formation of an imaginary potential well,
whose depth is expected, on the basis of the Dirac phenomenology analysis 
of the scattering of polarized intermediate-energy nucleons , to increase 
quadratically with energy above the threshold of pion production.

An important step in this direction is provided by Ref.~\cite{Typel:2002:AA},
where a relativistic optical potential suited to proton-nucleus
scattering at intermediate energies is derived in the frame of
a relativistic mean field model with density-dependent 
meson-nucleon couplings. The rearrangement contributions to
the nucleon self-energies, due to the density dependence of the
meson-nucleon couplings, are necessary in order to get a
thermodynamically consistent theory. These effects could be introduced
in the present model along the lines of Ref.~\cite{Niksic:2002:AAb},
where a new density-dependent parametrization, DD-ME1, has been
applied with success to the description of various properties 
of magic and semi-magic nuclei, and could lend itself to an
extension to open-shell nuclei without conceptual difficulties.

\newpage
\begin{table}
\caption{
 The parameter sets of the relativistic
 mean-field model.
	}
\begin{center}
\begin{tabular}{lrr}
Parameter   & NL1         & NL3 \\ 
\hline
$m_n$/MeV      & 938.000 & 939.0000 \\
$m_\sigma$/MeV & 492.250 & 508.1941 \\
$m_\omega$/MeV & 795.359 & 782.5010 \\
$m_\rho$/MeV   & 763.000 & 763.0000 \\
$g_\sigma$     &  10.138 &  10.2169 \\
$g_2$/$\mathrm{fm}^{-1}$ & -12.172  & -10.4307 \\
$g_3$          & -36.265 & -28.8851 \\
$g_\omega$     &  13.285 &  12.8675 \\
$g_\rho$       &   4.975 &   4.4744 \\
\end{tabular}
\end{center}
\label{tab:mf-par}
\end{table}
\begin{table}
\caption{ 
         Parameters of the linear energy dependence of the effective
	 nucleon mass, pairing strengths (from Refs.
	 \protect\cite{Tondeur:1979:AA,Cerf:1994:AAa}), and  
	 phonon energies ($\mathrm{E}(2^+)=30/\mathrm{A}^{2/3}$ MeV,
	 $\mathrm{E}(3^-)=50/\mathrm{A}^{2/3}$ MeV) used in this work.
	 For $^{208}$Pb the quadrupole and octupole phonon energies
         are the experimental energies of the $2^+_1$ and $3^+_1$
	 levels, respectively. The latter is the first excited level, and systematics
	 do not give the correct oder of levels.
	}
\begin{center}
\begin{tabular}{l|ccccc}
Nuclei       & $\alpha$(SE) & $G_n$ [MeV] & $G_p$ [MeV] 
                            & E($2^+$) [MeV] & E($3^-$) [MeV]\\ 
\hline
$^{208}$Pb   & -0.2  & 0.076   & 0.091	& 4.085	    & 2.615 \\
$^{114}$Sn   & -0.3  & 0.12   & 0.13	& 1.276     & 2.127 \\
$^{60}$Ni    & -0.1  & 0.21   & 0.19	& 1.957     & 3.262 \\
\end{tabular}
\end{center}
\label{tab:calc-par}
\end{table}
\begin{table}
\caption{ 
         Fits of the BBF (\protect\ref{eq:bbf}) and 
	 GBF (\protect\ref{eq:gbf}) parameters 
	 to the calculated total state densities.
	}
\begin{center}
\begin{tabular}{l|ccc|cccc}
& \multicolumn{3}{c|}{BBF} & \multicolumn{4}{c}{GBF} \\
Nuclei       & a [$\mathrm{MeV}^{-1}$] & $E_0$ [MeV] & $\chi^2$ 
             & a [$\mathrm{MeV}^{-1}$] & $E_0$ [MeV] & $\xi$ & $\chi^2$ \\ 
\hline
$^{208}$Pb   & 13.000 & 6.000    & 0.14  
             & 4.717  & 2.466    & 0.6354  & 0.015 \\
$^{114}$Sn   & 12.16  & -0.384   & 0.044 
             & 28.064 &  2.643 	 & 0.4226 & 0.01  \\
$^{60}$Ni    &  6.346 & -0.228   & 0.08 
             &  4.634 & -1.719   & 0.5410 & 0.08   \\
\end{tabular}
\end{center}
\label{tab:fit-par}
\end{table}
\end{document}